\documentclass{raa}

\usepackage{graphicx,times}
\usepackage{natbib}
\usepackage{amssymb,amsmath}
\bibpunct{(}{)}{;}{a}{}{,}

\usepackage[pagebackref=true]{hyperref}

\begin{document}

\title{Spins of SMBHs in distant low luminosity AGNs\\ {\it \small Accepted for publication in Research in Astronomy and Astrophysics.}}

   \setcounter{page}{1}

\author{M.Yu. Piotrovich\inst{}, S.D. Buliga\inst{}, T.M. Natsvlishvili\inst{}}

\institute{Central astronomical observatory at Pulkovo, St.-Petersburg, 196140, Russia;
{\it mpiotrovich@mail.ru}\\
\vs \no
}

\abstract{We estimated radiative efficiency, spin and SMBH mass values for sample of 33 distant low luminosity AGNs. The distribution of the estimated spin values (majority of objects have a spin greater than 0.8) is fairly typical for many types of AGNs. The dependence of the estimated spin values on the estimated SMBH masses shows strong correlation between them, which suggests a rapid increase in spin with mass, i.e. that the main mechanism of mass growth in this case is disk accretion. We did not find any significant qualitative differences in the spin characteristics between our objects and objects of other types considered in the paper.
}

   \authorrunning{M.Yu. Piotrovich et al. }            
   \titlerunning{Spins of SMBHs in distant low luminosity AGNs}  
   \maketitle

%
\section{Introduction}

Low-luminosity active galactic nuclei (LLAGN) represent a significant population of active galaxies, characterized by faint emission compared to classical quasars or Seyferts \citep{ptak01,Fernandez-Ontiveros12}. These faint systems challenge traditional selection techniques - optical emission lines become difficult to detect against host-galaxy light, and dust obscuration further hampers identification \citep{ho08a}. Building on the well-characterized population of LLAGNs in the nearby universe, James Webb Space Telescope's (JWST) near-infrared spectroscopy now enables us to extend LLAGN studies into the epoch of peak cosmic star formation and beyond \citep{merloni08}. Distant LLAGNs probe a regime of supermassive black hole (SMBH) growth that is largely unexplored at high redshift. Recent JWST observations from the JADES survey have uncovered 34 broad-line AGNs in the redshift range $1.5 < z < 9$ with bolometric luminosities down to $10^{43} \text{erg/s}$, revealing sub-Eddington accretion in low-mass host galaxies and hinting at an early emergence of local black hole - galaxy scaling relations \citep{juodzbalis25}.

In \citet{su17} authors showed that the spin (dimensionless rotation parameter) of the SMBH in LLAGNs can play a significant role in determining their physical properties and observational manifestations. In this work we decided to estimate the spin values of LLAGNs, perform a statistical analysis and compare the results with other types of AGNs.

\section{Analysis of initial sample}

We took the initial sample of 34 distant LLAGNs from \citet{juodzbalis25}. Based on the results of the initial analysis, we decided to exclude the object GS-179198 from consideration due to very large errors in determining the SMBH mass and bolometric luminosity. The list of objects and their physical parameters can be seen in Table~\ref{table_01}.

\begin{figure}[ht!]
   \centering
   \includegraphics[bb= 60 25 735 530, clip, width=0.7\textwidth]{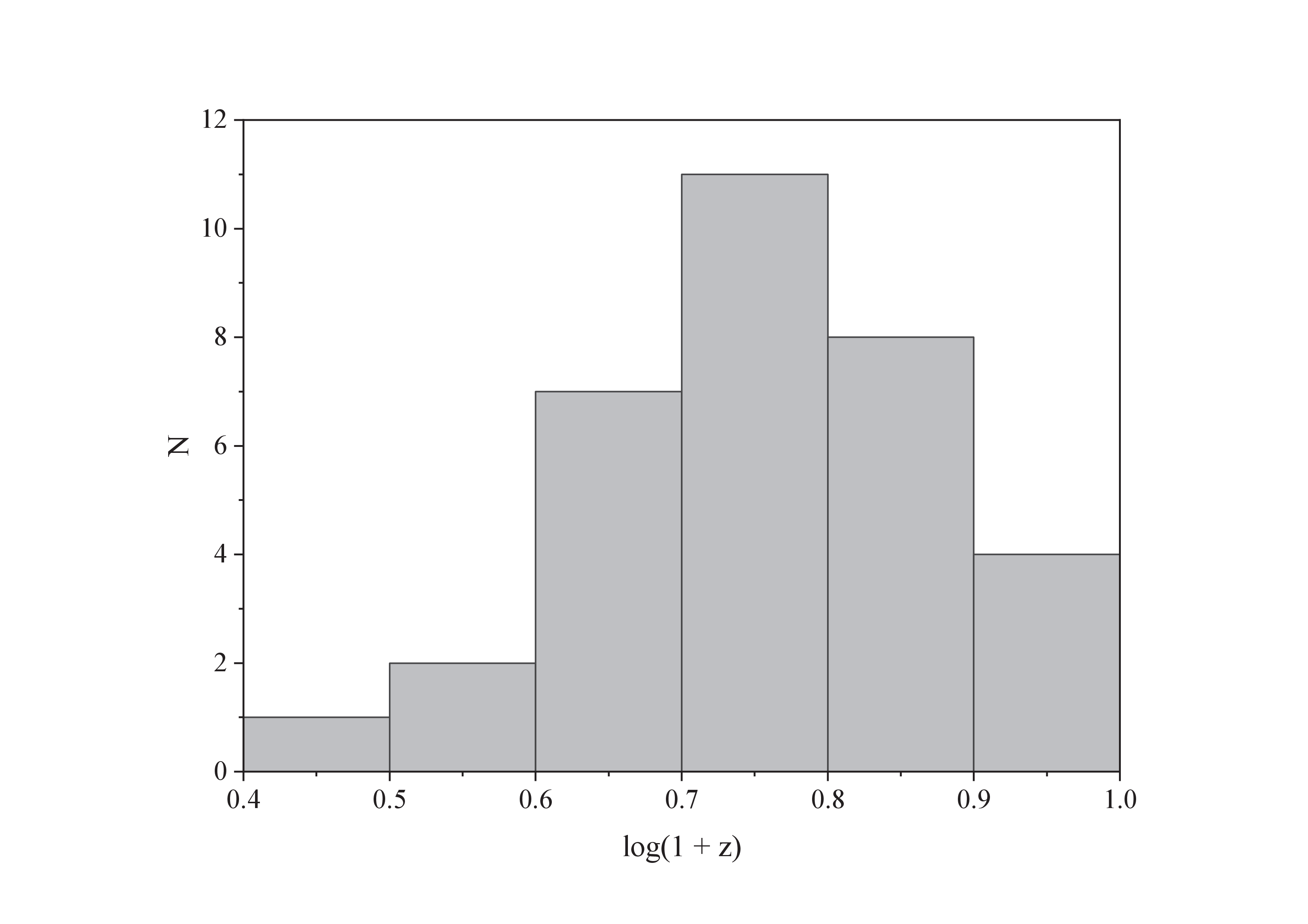}
   \caption{Distribution of the cosmological redshift for the initial sample.}
   \label{hist_z}
\end{figure}

We can see in Fig.~\ref{hist_z} the distribution of the cosmological redshift for the initial sample. The distribution appears to be close to log-normal, so we can conclude that there are no significant differences from a random distribution.

\begin{figure}[ht!]
   \centering
   \includegraphics[bb= 60 25 735 530, clip, width=0.7\textwidth]{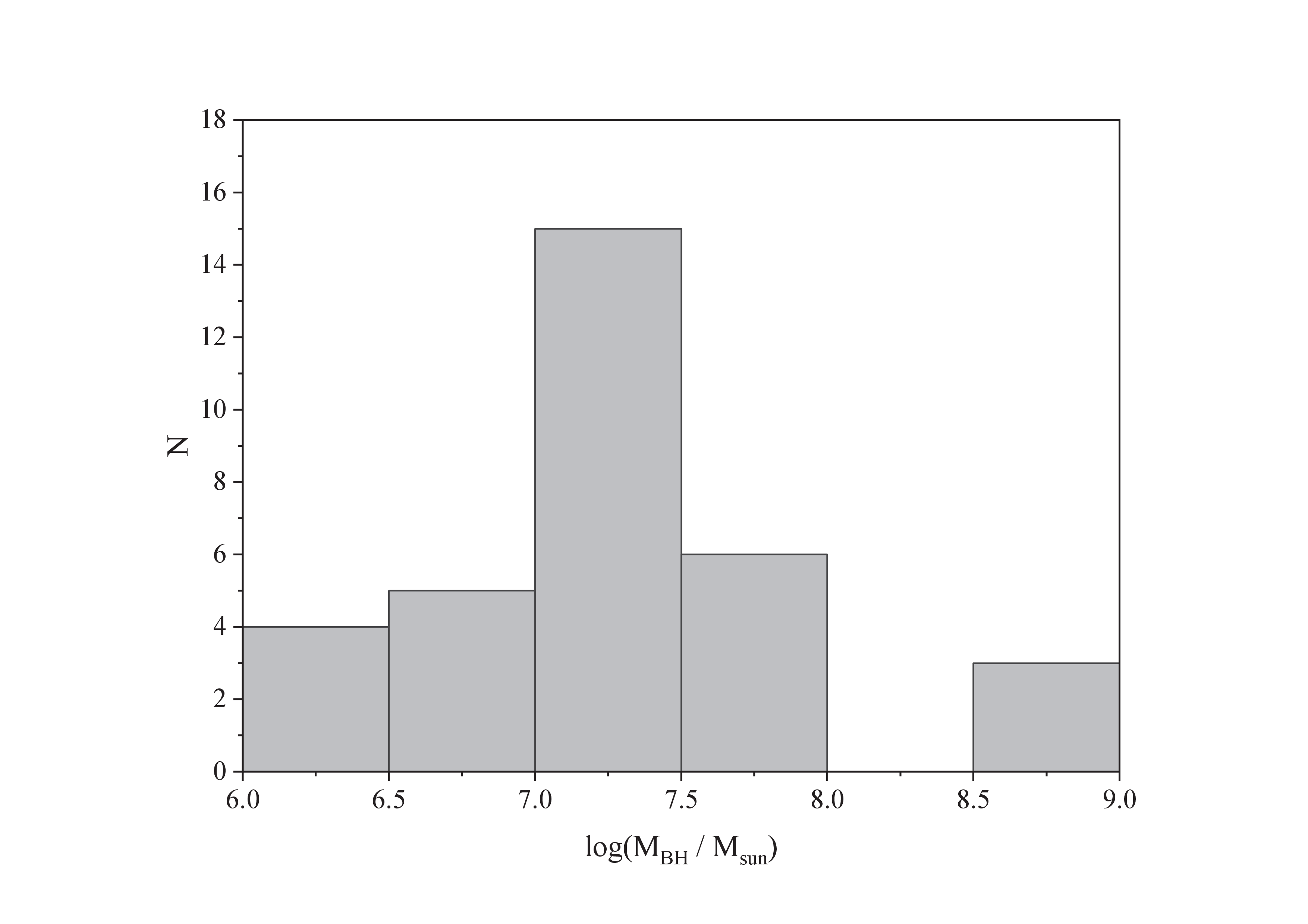}
   \caption{Distribution of the SMBH mass for the initial sample.}
   \label{hist_M}
\end{figure}

Fig.~\ref{hist_M} shows the distribution of the SMBH mass for the initial sample. This distribution also looks somewhat similar to log-normal, but has a pronounced peak in the area of $7.0 < \log(M_\text{BH} / M_\odot) < 7.5$ which is most likely due to the method of selecting objects in \citet{juodzbalis25}.

\begin{figure}[ht!]
   \centering
   \includegraphics[bb= 60 25 740 530, clip, width=0.7\textwidth]{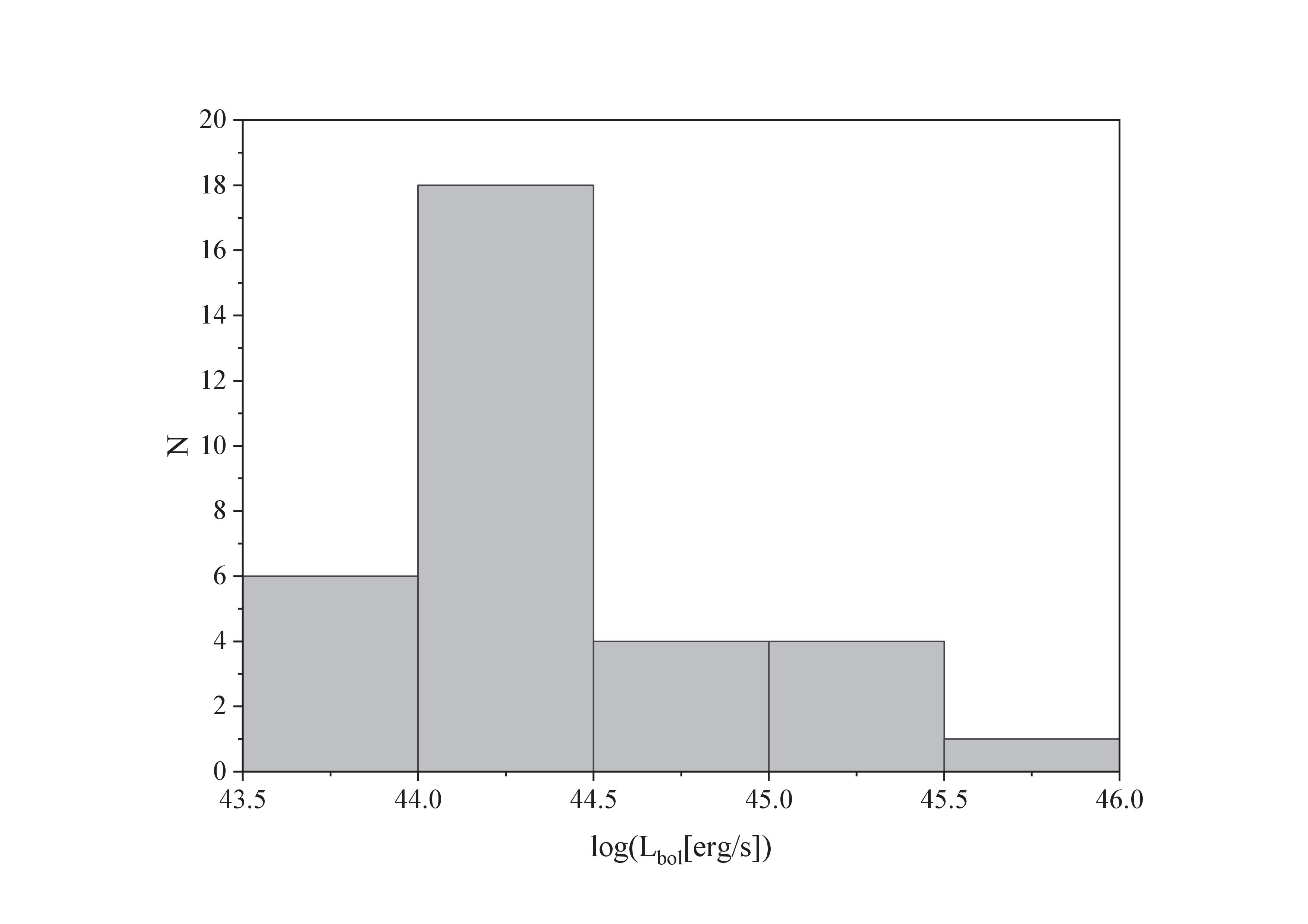}
   \caption{Distribution of the bolometric luminosity for the initial sample.}
   \label{hist_L_bol}
\end{figure}

Fig.~\ref{hist_L_bol} demonstrates the distribution of the bolometric luminosity for the initial sample. It has a pronounced peak in the area of $44.0 < \log(L_\text{bol}[\text{erg/s}]) < 44.5$. This is apparently due to the nature of the mass distribution.

\begin{figure}[ht!]
   \centering
   \includegraphics[bb= 60 25 740 530, clip, width=0.7\textwidth]{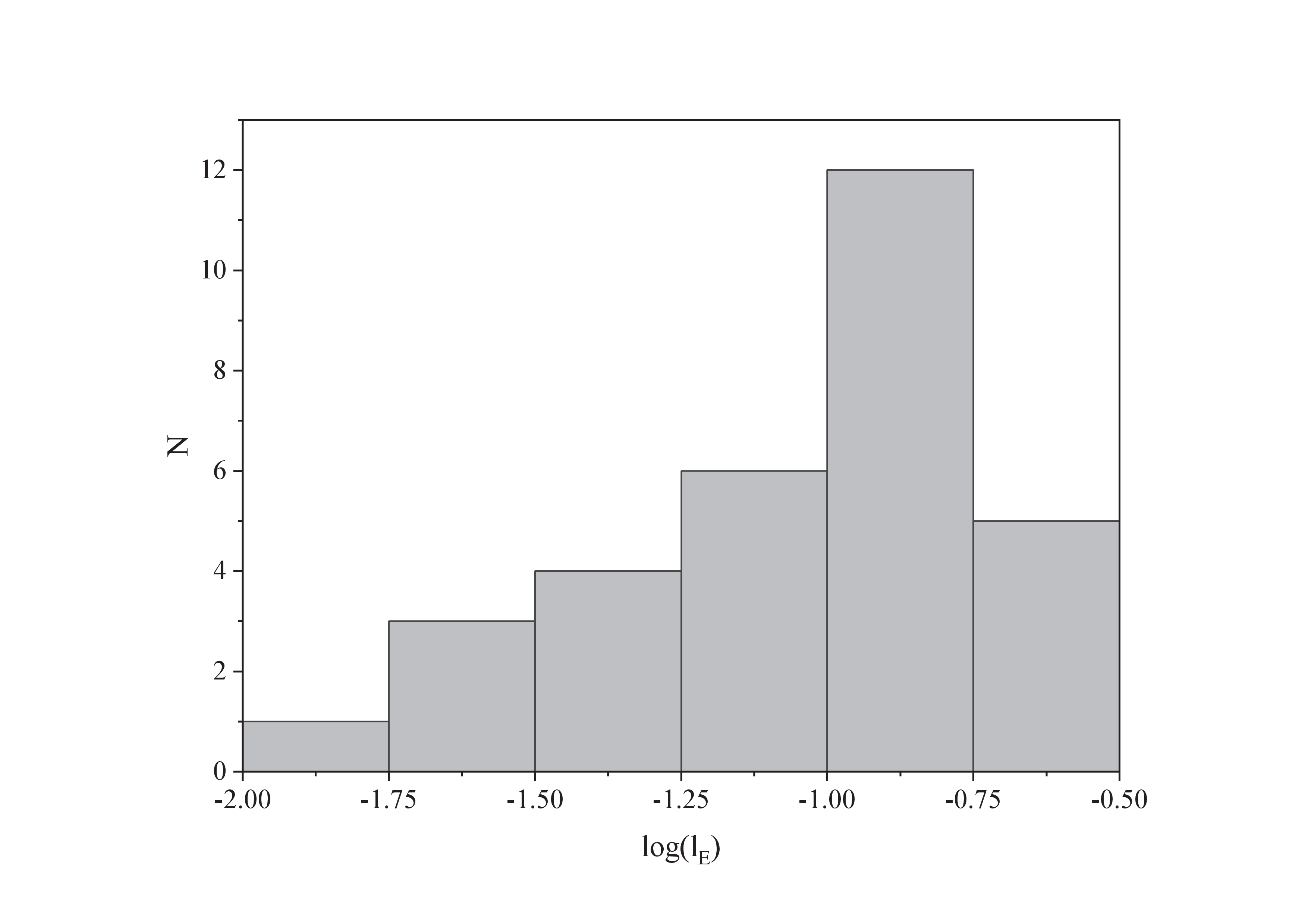}
   \caption{Distribution of the Eddington ratio for the initial sample.}
   \label{hist_l_E}
\end{figure}

In Fig.~\ref{hist_l_E} we can see the distribution of the Eddington ratio $l_E = L_\text{bol} / L_\text{Edd}$ (where $L_\text{Edd}$ is the Eddington luminosity) for the initial sample. All objects have sub-Eddington luminosity with a peak in the region of $-1.0 < \log(l_\text{E}) < -0.75$. It can be assumed that for objects of this type (distant LLAGNs) this is the most typical value.

\begin{figure}[ht!]
   \centering
   \includegraphics[bb= 80 25 720 530, clip, width=0.7\textwidth]{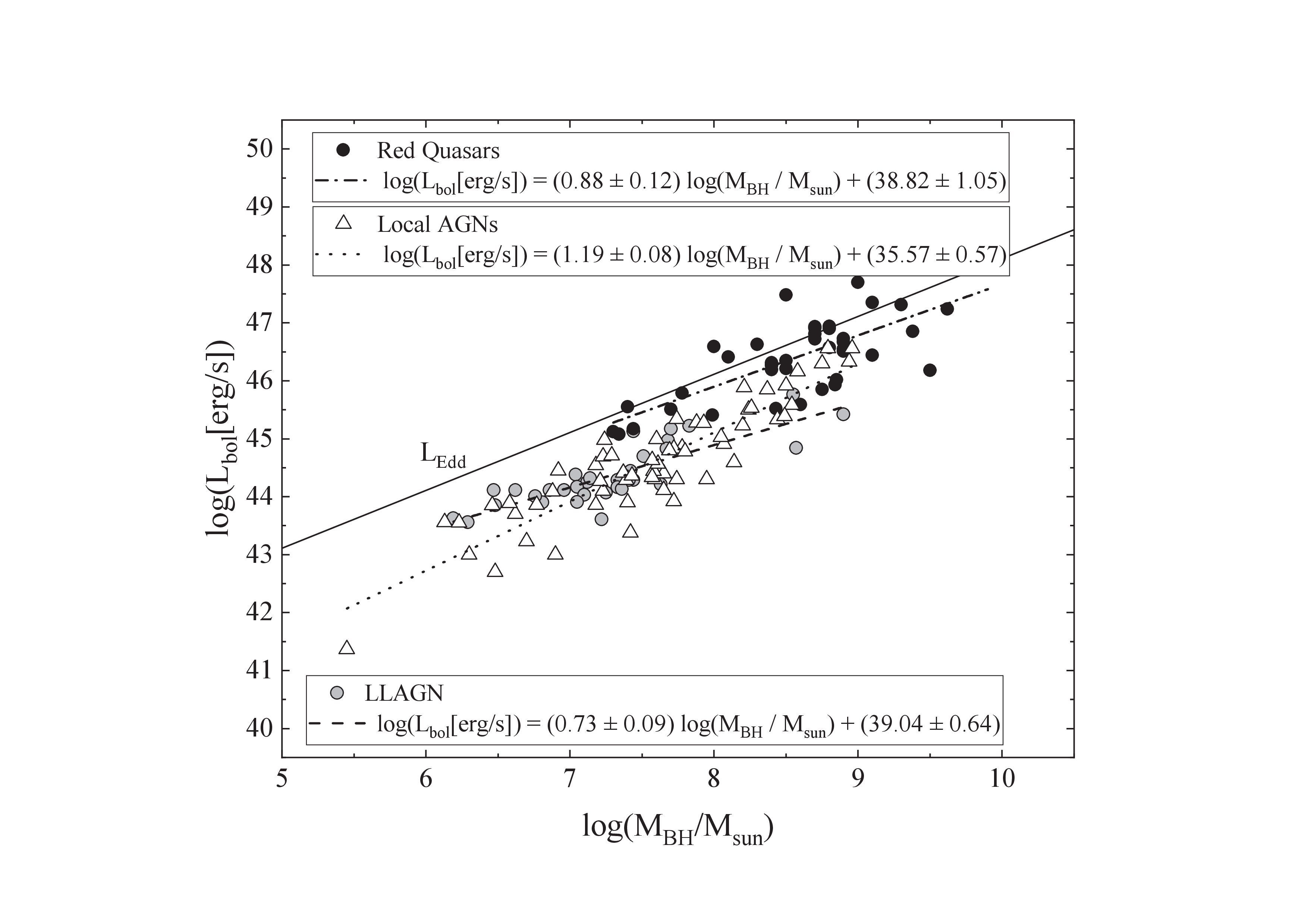}
   \caption{Dependence of the bolometric luminosity on the SMBH mass for the initial sample, for sample of red quasars from \citet{piotrovich24} and for sample of local AGNs from \citet{piotrovich22}.}
   \label{M-L_bol}
\end{figure}

And finally, Fig.~\ref{M-L_bol} shows the dependence of the bolometric luminosity on the SMBH mass for the initial sample. Solid line represents Eddington luminosity as a function of SMBH mass $\log(L_\text{Edd}) = \log(M_\text{BH} / M_\odot) + 38.11$. Bolometric luminosity and SMBH mass show strong correlation (Pearson correlation coefficient is 0.83). Linear fitting gives us: $\log(L_\text{bol}\text{[erg/s]}) = (0.73 \pm 0.09) \log(M_\text{BH} / M_\odot)) + (39.04 \pm 0.64)$. It can be noted that a similar dependence, for example, for red quasars, obtained by us in \citet{piotrovich24} shows a little stronger dependence of luminosity on mass in the form of $\log(L_\text{bol}\text{[erg/s]}) = (0.88 \pm 0.12) \log(M_\text{BH} / M_\odot)) + (38.82 \pm 1.05)$. For comparison, we also provide data for local AGNs (mostly Seyfert galaxies) from \citet{piotrovich22} for which linear fitting gives us $log(L_\text{bol}\text{[erg/s]}) = (1.19 \pm 0.08) \log(M_\text{BH} / M_\odot)) + (35.57 \pm 0.57)$.


\section{Method for estimating spins}

\begin{figure}[ht!]
\centering
\includegraphics[bb= 55 35 700 535, clip, width= 0.7\columnwidth]{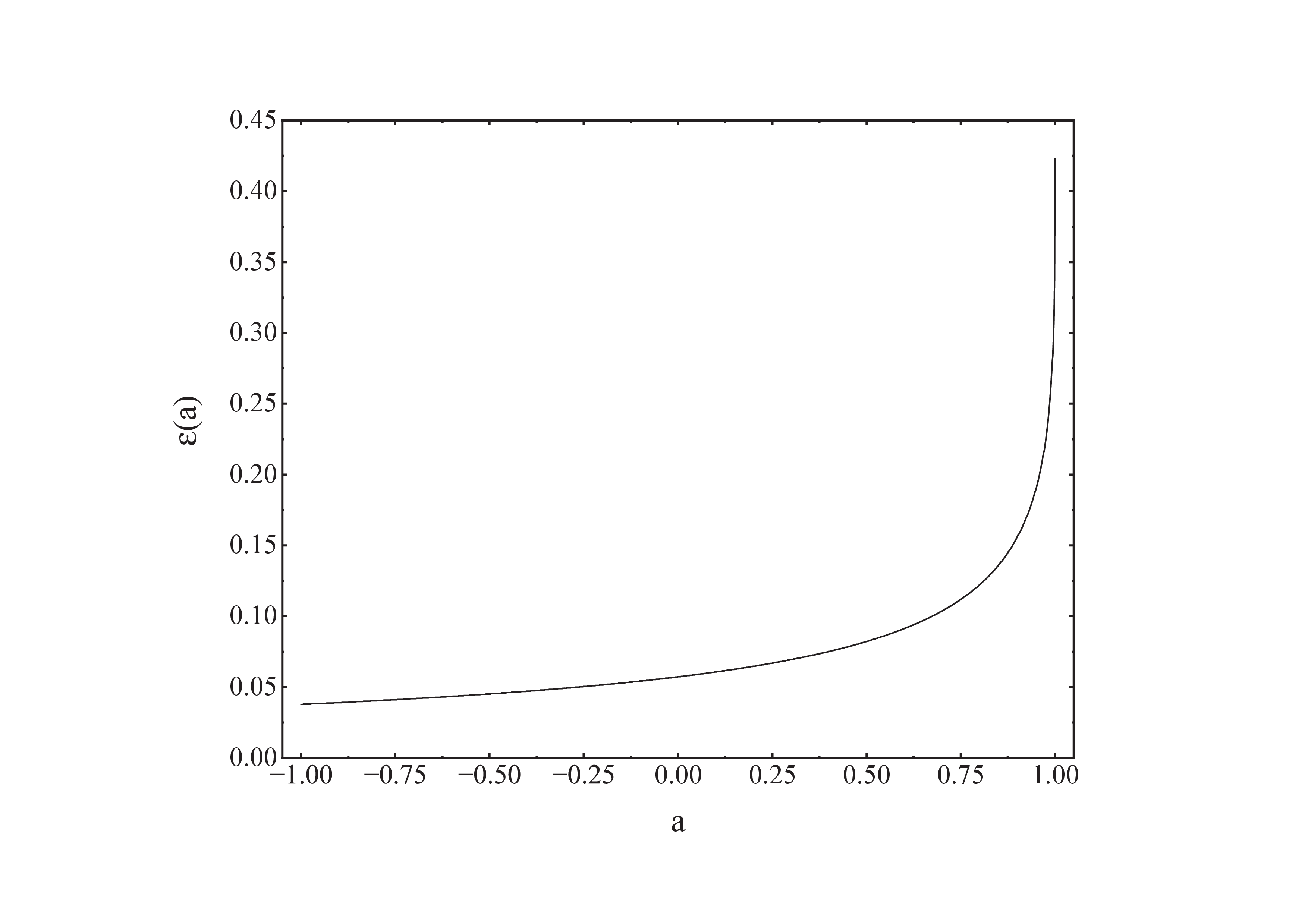}
\caption{Radiative efficiency coefficient as the function of the spin value.}
  \label{a_eps}
\end{figure}

The spin of a black hole can be estimated by evaluating the radiative efficiency $\varepsilon(a)$ of its accretion disk, which is directly influenced by the black hole’s spin value \citep{bardeen72,novikov73,krolik07,krolik07b} (see Fig.~\ref{a_eps}). Radiative efficiency is defined as $\varepsilon = L_\text{bol} / (\dot{M} c^2)$, where $L_\text{bol}$ represents the bolometric luminosity of the AGN or quasar, and $\dot{M}$ is the mass accretion rate. The efficiency is expected to lie within the range $0.039 < \varepsilon(a) < 0.324$, corresponding to spin values of $-1.0 < a \leq 0.998$ \citep{thorne74}. Negative spin values indicate ''retrograde'' rotation, where the accretion disk and the black hole rotate in opposite directions.

Radiative efficiency is influenced by several factors, including the mass of the SMBH ($M_\text{BH}$), the inclination angle $i$ between the observer’s line of sight and the normal to the accretion disk, and the bolometric luminosity ($L_\text{bol}$) \citep{davis11,raimundo11,du14,trakhtenbrot14,lawther17}. These estimates are based on statistical analyses of observational data and rely on the Shakura–Sunyaev accretion disk model \citep{shakura73}. It should be noted that Shakura-Sunyaev model is not ideal and has shortcomings. However at the moment, generally speaking, there is no better model and most authors use it. Thus, we use the Shakura-Sunyaev model in order to be able to use results and methods of other authors and to compare our results with the results of other authors (and our previous results). Given that we are studying extremely distant objects, we adopt the model presented in \citet{trakhtenbrot14}, which was specifically developed for such cases:

\begin{equation}
 \varepsilon \left( a \right) =  0.073 \left(\frac{L_\text{bol}}{10^{46}\text{erg/s}}\right) \left(\frac{L_\text{opt}}{10^{45}\text{erg/s}}  \right)^{-1.5} \times \left(\frac{4400\text{\AA}}{5100\text{\AA}}\right)^{-2} M_8 \mu^{1.5},
 \label{eq03}
\end{equation}

\noindent where $L_\text{opt}$ is the optical luminosity (at 4400\AA), $M_8 = \frac{M_\text{BH}}{10^8 M_{\odot}}$ and $\mu = \cos{(i)}$.

Determining bolometric correction factors—used to derive the luminosity at a specific wavelength from the bolometric luminosity—is a complex task. Values reported in the literature can vary by a factor of 2–3 \citep{richards06,hopkins07,cheng19,netzer19,duras20}. The use of bolometric corrections is not a very accurate method. However, it is not possible to apply more accurate methods to our current objects. In this study, we adopt for consistency with our previous work the definition of optical luminosity $L_\text{opt}$ from \citet{hopkins07}:
\begin{equation}
  \frac{L_\text{bol}}{L_\text{opt}} = 6.25 \left(\frac{L_\text{bol}}{10^{10} L_\odot}\right)^{-0.37} + 9.0 \left(\frac{L_\text{bol}}{10^{10} L_\odot}\right)^{-0.012},
\end{equation}

\noindent where $L_\odot$ is the Sun luminosity.

Accurately determining the inclination angle $i$ from observational data remains a challenging and unresolved problem. It is common practice to assume a fixed average value for $i$ (usually around 30-40 degrees). In this work, however, we adopted a more adaptive approach: for each object, we initially assumed an average inclination of $i = 30^{\circ}$. If this value did not yield a physically meaningful result through our numerical method, we systematically adjusted the angle in increments of $5^{\circ}$—both lower and higher—until a valid solution was obtained.

The estimated mass of a SMBH depends on the inclination angle $i$, as the commonly used mass determination methods rely on the relation \citep{decarli08}:
\begin{equation}
  M_\text{BH} = \frac{R_\text{BLR} V_\text{BLR}^2}{G},
  \label{eq_M_BH}
\end{equation}

\noindent where $R_\text{BLR}$ is the characteristic radius of the broad-line region (BLR) in the accretion disk, $V_\text{BLR}$ is the typical velocity of accreting matter within the BLR, and $G$ is the gravitational constant. The velocity $V_\text{BLR}$ can be inferred from observations—for instance, by measuring the full width at half maximum (FWHM) of the H$\beta$ spectral line:

\begin{equation}
  V_\text{BLR} = f \times FWHM(\text{H}\beta).
  \label{eq_V_BLR}
\end{equation}

\noindent $f$ is coefficient related to the geometry of the accretion disk that equal to \citep{decarli08}:
\begin{equation}
  f = \left(2\sqrt{\left(\frac{H}{R}\right)^2 + \sin^2{i}}\right)^{-1}.
  \label{eq_f}
\end{equation}

\noindent Here $H/R$ is the ratio of the thickness to the radius of the disk. For our objects we assume geometrically thin disks, so $H/R \ll 1$ and
\begin{equation}
  f \approx \frac{1}{2 \sin{i}}.
  \label{eq_f1}
\end{equation}

We can estimate $R_\text{BLR}$ using method from \citet{bentz13}:
\begin{equation}
  \log(R_\text{BLR} / 1\, \text{lt-day}) \approx 1.5 + 0.5\log(L_{5100} / 10^{44} \text{erg/s}),
  \label{eq_R_BLR}
\end{equation}

\noindent where $L_\text{5100}$ is luminosity at 5100\AA. We used bolometric correction for luminosity at 5100\AA\, from \citet{richards06}: $\log(L_\text{bol}\text{[erg/s]}) = \log(L_\text{5100}\text{[erg/s]}) + 1.01$.

In the process of estimating radiative efficiencies for different values of inclination angle we for self-consistency estimated new mass values $M^*_\text{BH}$ (Eq.(\ref{eq_M_BH})), considering that masses from \citet{juodzbalis25} was obtained for $i \approx 30^{\circ}$.

The spins were estimated numerically using the technique from \citet{bardeen72} {(see Fig.\ref{a_eps})}:
\begin{equation}
  \varepsilon(a) = 1 - \frac{R_\text{ISCO}^{3/2} - 2 R_\text{ISCO}^{1/2} + |a|}{R_\text{ISCO}^ {3/4}\left(R_\text{ISCO}^{3/2} - 3 R_\text{ISCO}^{1/2} + 2 |a|\right)^{1/2}}.
  \label{eq01}
\end{equation}

\noindent Here $R_\text{ISCO}$ is the radius of the innermost stable circular orbit that depends on the spin value:
\begin{equation}
  \begin{array}{l}
   R_\text{ISCO}(a) = \\
   = 3 + Z_2 \pm [(3 - Z_1)(3 + Z_1 + 2 Z_2)]^{1/2},\\
   Z_1 = 1 + (1 - a^2)^{1/3}\left[(1 + a)^{1/3} + (1 - a)^{1/3}\right],\\
   Z_2 = (3 a^2 + Z_1^2)^{1/2},
  \end{array}
  \label{eq02}
\end{equation}

\noindent where ''-'' is refers to $a \geq 0$, and ''+'' to $a < 0$.


\section{Analysis of the obtained data}

\begin{table}[ht!]
\begin{center}
\caption[]{Table shows object name, cosmological redshift $z$, SMBH mass $M_\text{BH}$ and bolometric luminosity $L_\text{bol}$ (in erg/s) from \citet{juodzbalis25} and results of our estimations: inclination angle $i$ (in degrees), SMBH mass $M^*_\text{BH}$, radiative efficiency $\varepsilon$ and spin value $a$.}
\begin{tabular}{lccccccc}
\hline\noalign{\smallskip}
Object & $z$ & $\log{\frac{M_\text{BH}}{M_\odot}}$ & $\log{L_\text{bol}}$ & $i$ & $\log{\frac{M^*_\text{BH}}{M_\odot}}$ & $\varepsilon$ & $a$\\
\hline\noalign{\smallskip}
GS-30148179 & 5.922 & 7.12 & 44.25 & 30 & 7.12 & 0.108 &  0.736\\
GS-10013704 & 5.919 & 7.44 & 44.29 & 30 & 7.44 & 0.211 &  0.970\\
  GS-210600 & 6.306 & 7.41 & 44.29 & 30 & 7.41 & 0.197 &  0.958\\
  GS-209777 & 3.709 & 8.90 & 45.42 & 55 & 8.47 & 0.241 &  0.984\\
  GS-204851 & 5.480 & 7.68 & 44.97 & 30 & 7.68 & 0.135 &  0.852\\
  GS-172975 & 4.741 & 7.25 & 44.07 & 30 & 7.25 & 0.192 &  0.954\\
  GS-159717 & 5.077 & 7.44 & 45.13 & 30 & 7.44 & 0.062 &  0.162\\
  GS-159438 & 3.239 & 6.47 & 44.11 & 25 & 6.62 & 0.045 & -0.452\\
   GN-77652 & 5.229 & 6.62 & 44.11 & 30 & 6.62 & 0.042 & -0.624\\
   GN-73488 & 4.133 & 7.83 & 45.22 & 30 & 7.83 & 0.134 &  0.848\\
   GN-62309 & 5.172 & 6.29 & 43.56 & 30 & 6.29 & 0.049 & -0.262\\
   GN-61888 & 5.874 & 7.04 & 44.38 & 30 & 7.04 & 0.073 &  0.384\\
   GN-53757 & 4.447 & 7.33 & 44.29 & 30 & 7.33 & 0.164 &  0.918\\
   GS-49729 & 3.189 & 7.67 & 44.83 & 30 & 7.67 & 0.161 &  0.912\\
   GS-38562 & 4.822 & 7.51 & 44.70 & 30 & 7.51 & 0.134 &  0.848\\
   GN-38509 & 6.678 & 8.57 & 44.84 & 55 & 8.14 & 0.253 &  0.988\\
   GN-29648 & 2.960 & 6.81 & 43.90 & 30 & 6.81 & 0.092 &  0.618\\
   GN-28074 & 2.259 & 8.55 & 45.76 & 35 & 8.43 & 0.238 &  0.984\\
   GN-20621 & 4.682 & 7.05 & 44.17 & 30 & 7.05 & 0.104 &  0.710\\
   GS-17341 & 3.598 & 6.76 & 44.01 & 30 & 6.76 & 0.068 &  0.294\\
   GS-13329 & 3.936 & 6.86 & 44.11 & 30 & 6.86 & 0.074 &  0.400\\
   GN-11836 & 4.409 & 6.96 & 44.11 & 30 & 6.96 & 0.093 &  0.626\\
    GS-9598 & 3.324 & 6.48 & 43.85 & 30 & 6.48 & 0.047 & -0.354\\
    GS-8083 & 4.753 & 7.10 & 44.03 & 30 & 7.10 & 0.145 &  0.880\\
    GN-2916 & 3.664 & 7.05 & 43.91 & 30 & 7.05 & 0.157 &  0.906\\
    GN-1093 & 5.594 & 7.14 & 44.32 & 30 & 7.14 & 0.101 &  0.690\\
     GN-954 & 6.759 & 7.70 & 45.17 & 30 & 7.70 & 0.107 &  0.730\\
  GS-200679 & 4.547 & 6.19 & 43.63 & 25 & 6.34 & 0.052 & -0.142\\
   GN-23924 & 1.676 & 7.22 & 43.61 & 35 & 7.10 & 0.267 &  0.992\\
GS-20057765 & 8.913 & 7.33 & 44.16 & 30 & 7.33 & 0.201 &  0.962\\
GS-20030333 & 7.891 & 7.42 & 44.44 & 30 & 7.42 & 0.161 &  0.912\\
  GS-164055 & 7.397 & 7.63 & 44.21 & 35 & 7.51 & 0.259 &  0.990\\
    GN-4685 & 7.415 & 7.36 & 44.13 & 30 & 7.36 & 0.225 &  0.978\\
\noalign{\smallskip}\hline
\end{tabular} \label{table_01}
\end{center}
\end{table}

We estimated radiative efficiency, spin and new SMBH mass values for our sample of objects. The results are presented at Table~\ref{table_01}. It should be noted that our estimates cannot be considered as exact values for each individual object. Rather, these results have statistical significance for the entire sample of objects as a whole. Also inclination angle values in this case should be considered as a lower estimate.

\begin{figure}[ht!]
   \centering
   \includegraphics[bb= 50 25 740 530, clip, width=0.7\textwidth]{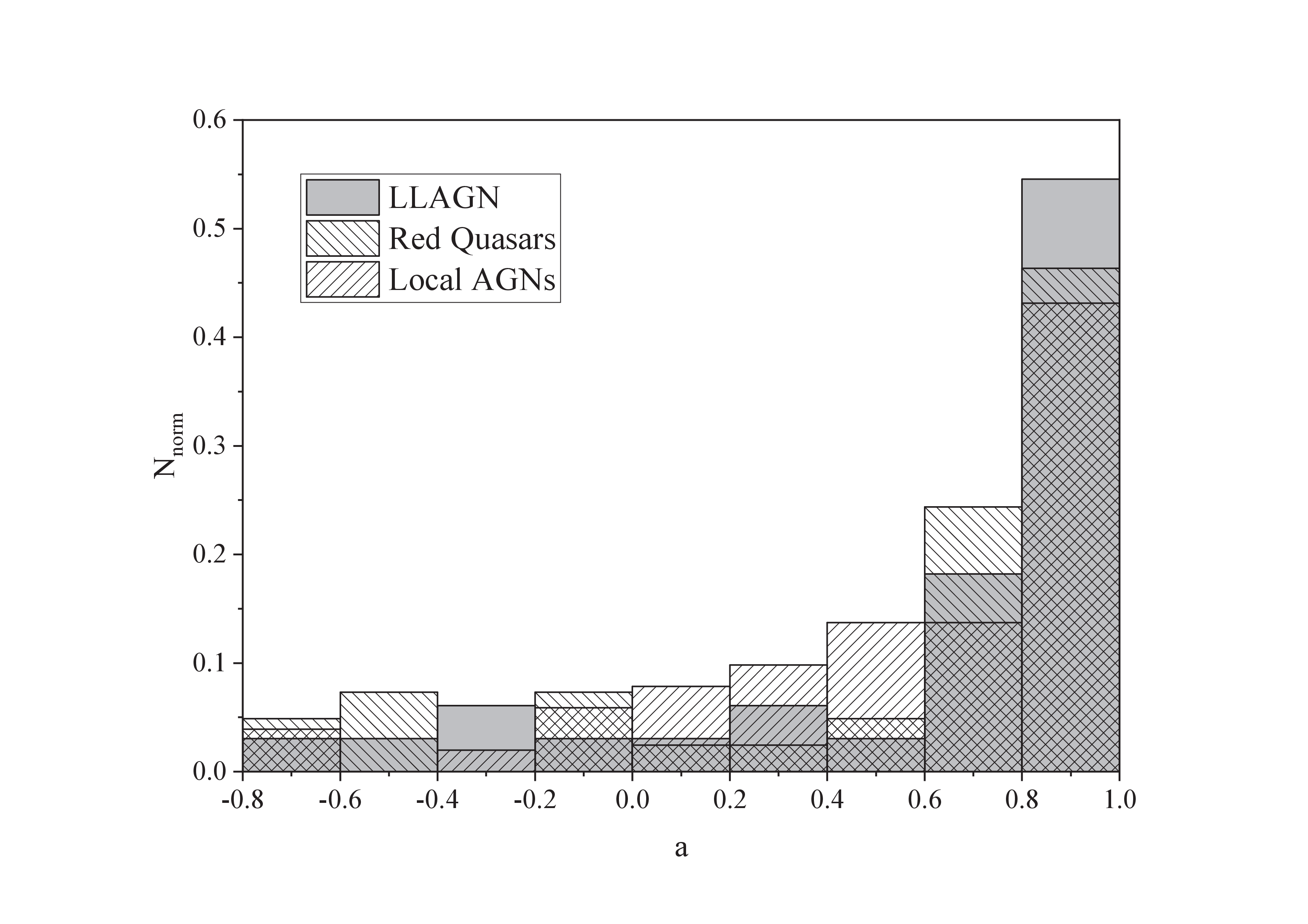}
   \caption{Normalized distribution of the estimated spin values for our sample, for sample of red quasars from \citet{piotrovich24} and for sample of local AGNs from \citet{piotrovich22}.}
   \label{hist_a}
\end{figure}

Fig.~\ref{hist_a} shows the normalized distribution of the estimated spin values. It can be seen that the majority of objects have a spin greater than 0.8. This distribution pattern is fairly typical for many types of AGNs: distant quasars \cite{trakhtenbrot14}, Seyfert galaxies \cite{afanasiev18,piotrovich22}, red quasars \cite{piotrovich24} and other types \citep{daly19,reynolds21,azadi23}. From this we can suggest that our objects in this regard apparently do not have significant differences from most objects of other types.

\begin{figure}[ht!]
   \centering
   \includegraphics[bb= 60 25 740 530, clip, width=0.7\textwidth]{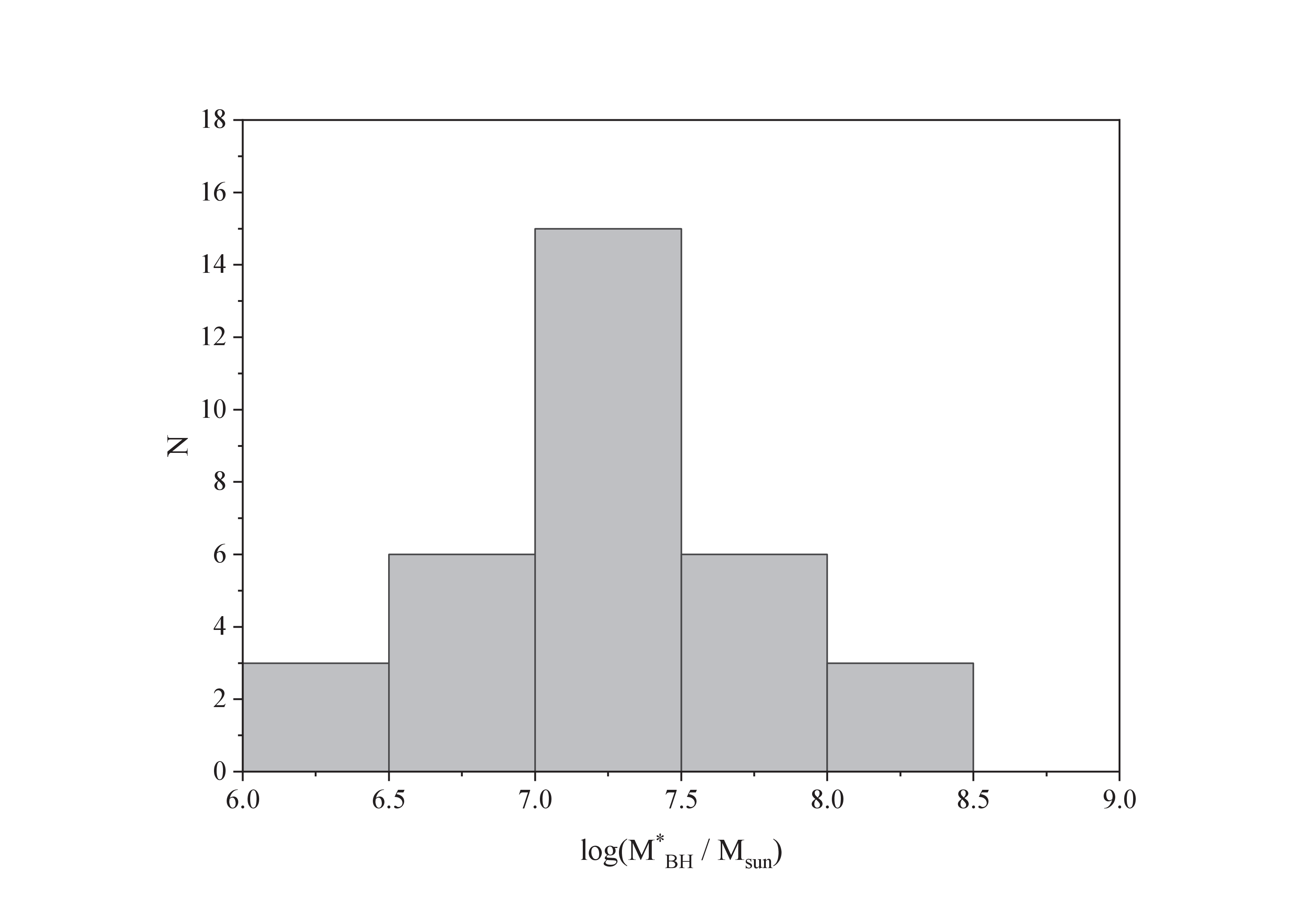}
   \caption{Distribution of the estimated SMBH masses.}
   \label{hist_M_star}
\end{figure}

Fig.~\ref{hist_M_star} presents the distribution of the new estimated SMBH masses. This distribution is practically no different from the initial one (Fig.~\ref{hist_M}), since for most objects our method gave a physically meaningful result at the standard inclination angle of 30 degrees. However, that does not mean that these objects actually have these exact inclination angles. The most that can be said is that the real angles do not differ much from 30 degrees, which is generally expected for objects of this type.

\begin{figure}[ht!]
   \centering
   \includegraphics[bb= 85 25 720 530, clip, width=0.7\textwidth]{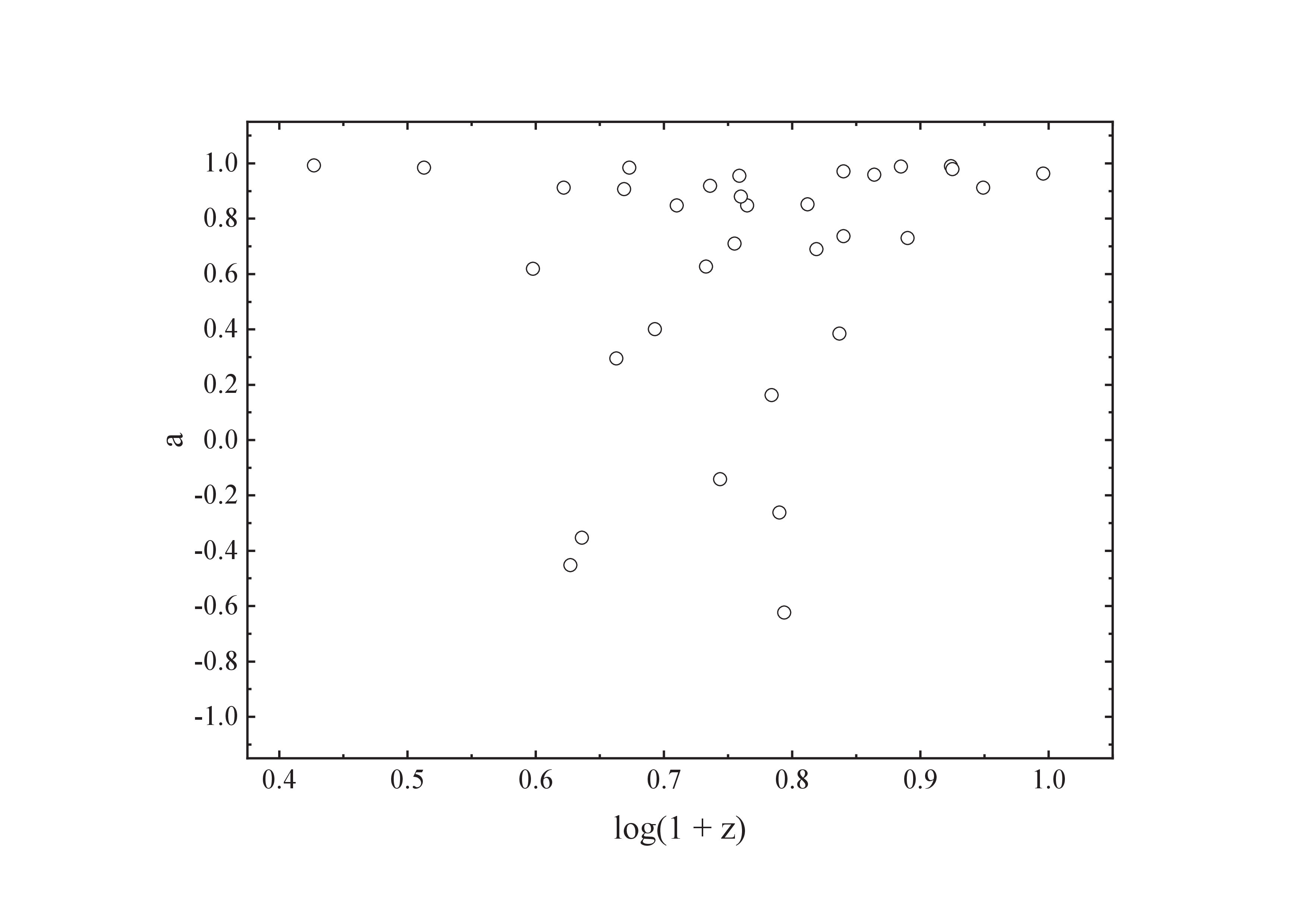}
   \caption{Dependence of the estimated spin values on the cosmological redshift.}
   \label{z-a}
\end{figure}

On Fig.~\ref{z-a} we can see the dependence of the estimated spin values on the cosmological redshift. In this case, there is no noticeable correlation between spin and redshift (Pearson correlation coefficient is 0.16). Thus, we do not observe a noticeable increase in spin over time, which is observed, for example, for ultra-luminous distant AGN \citep{piotrovich25}. It can be assumed that in this case it is connected with the selection bias in the initial sample. This is due to the fact that, firstly, the set of our objects is quite small and, secondly, all our objects are very distant and have a fairly small range of redshift values.

\begin{figure}[ht!]
   \centering
   \includegraphics[bb= 60 25 720 530, clip, width=0.7\textwidth]{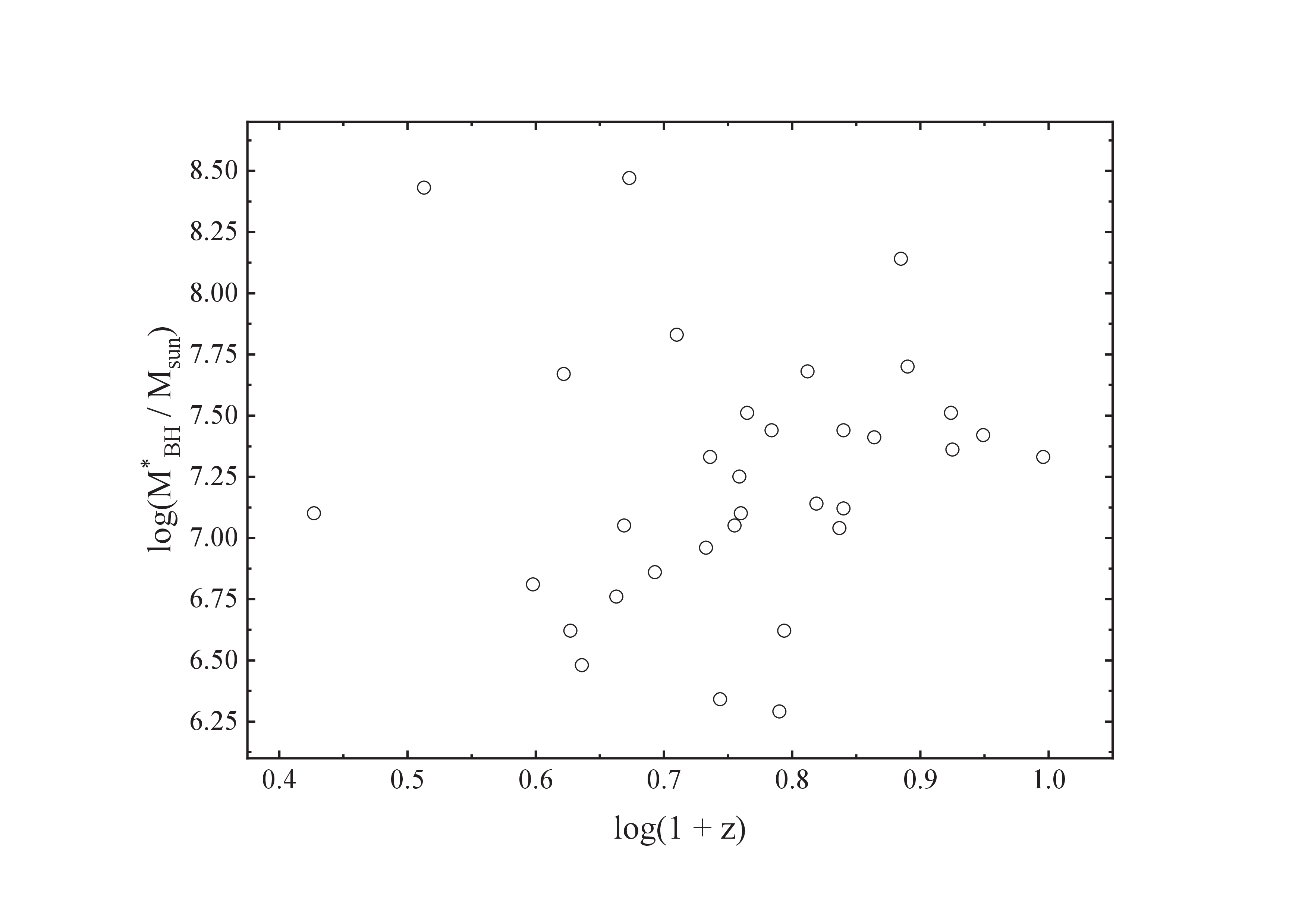}
   \caption{Dependence of the estimated SMBH masses on the cosmological redshift.}
   \label{z-M_star}
\end{figure}

Fig.~\ref{z-M_star} shows the dependence of the estimated SMBH masses on the cosmological redshift. There is also no noticeable correlation between SMBH mass and redshift (Pearson correlation coefficient is 0.09). A similar dependence for ultraluminous AGNs gave a noticeable anti-correlation \citep{piotrovich25} (i.e. the SMBHs masses grow noticeably over time). Apparently, this is also related to the selection bias, since in the original sample, objects were selected based on their low luminosity.

\begin{figure}[ht!]
   \centering
   \includegraphics[bb= 60 25 720 530, clip, width=0.7\textwidth]{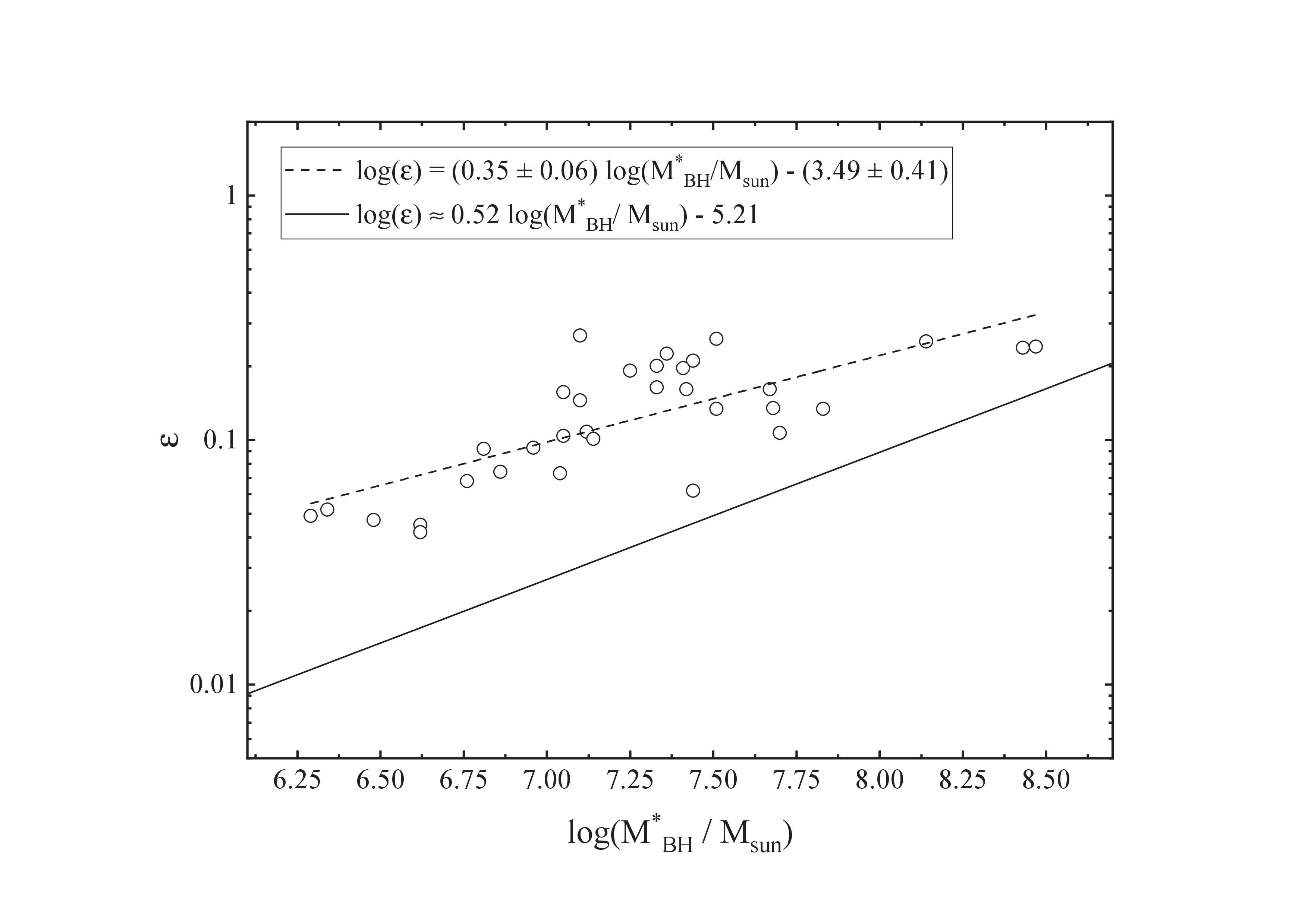}
   \caption{Dependence of the estimated radiative efficiency on the estimated SMBH masses. Solid line is linear fit for sample from \citet{davis11}.}
   \label{M_star-eps}
\end{figure}

Fig.~\ref{M_star-eps} presents the dependence of the estimated radiative efficiency on the estimated SMBH masses. There is strong correlation between radiative efficiency and SMBH masses (Pearson correlation coefficient is 0.75). Linear fitting gives us $\log(\varepsilon) = (0.35 \pm 0.06) \log(M^*_\text{BH} / M_\odot) - (3.49 \pm 0.41)$. We can compare this result with the result from \citet{davis11} (see Fig.9 from their paper). Written in a similar way to our result, it looks like this: $\log(\varepsilon) \approx 0.52 \log(M^*_\text{BH} / M_\odot) - 5.21$. In \citet{davis11} the objects have a much wider range of radiative efficiency values than our objects. There are objects with both larger and smaller values. As a result the slope of the linear fit is larger, but it is located below ours. In our set of objects, there are no small values of radiative efficiency, most likely due to the selection bias. These are extremely distant objects, and at very small values of radiative efficiency, they become too weak. There are also no large values in our sample, since low luminosity objects were initially selected.

\begin{figure}[ht!]
   \centering
   \includegraphics[bb= 60 25 720 530, clip, width=0.7\textwidth]{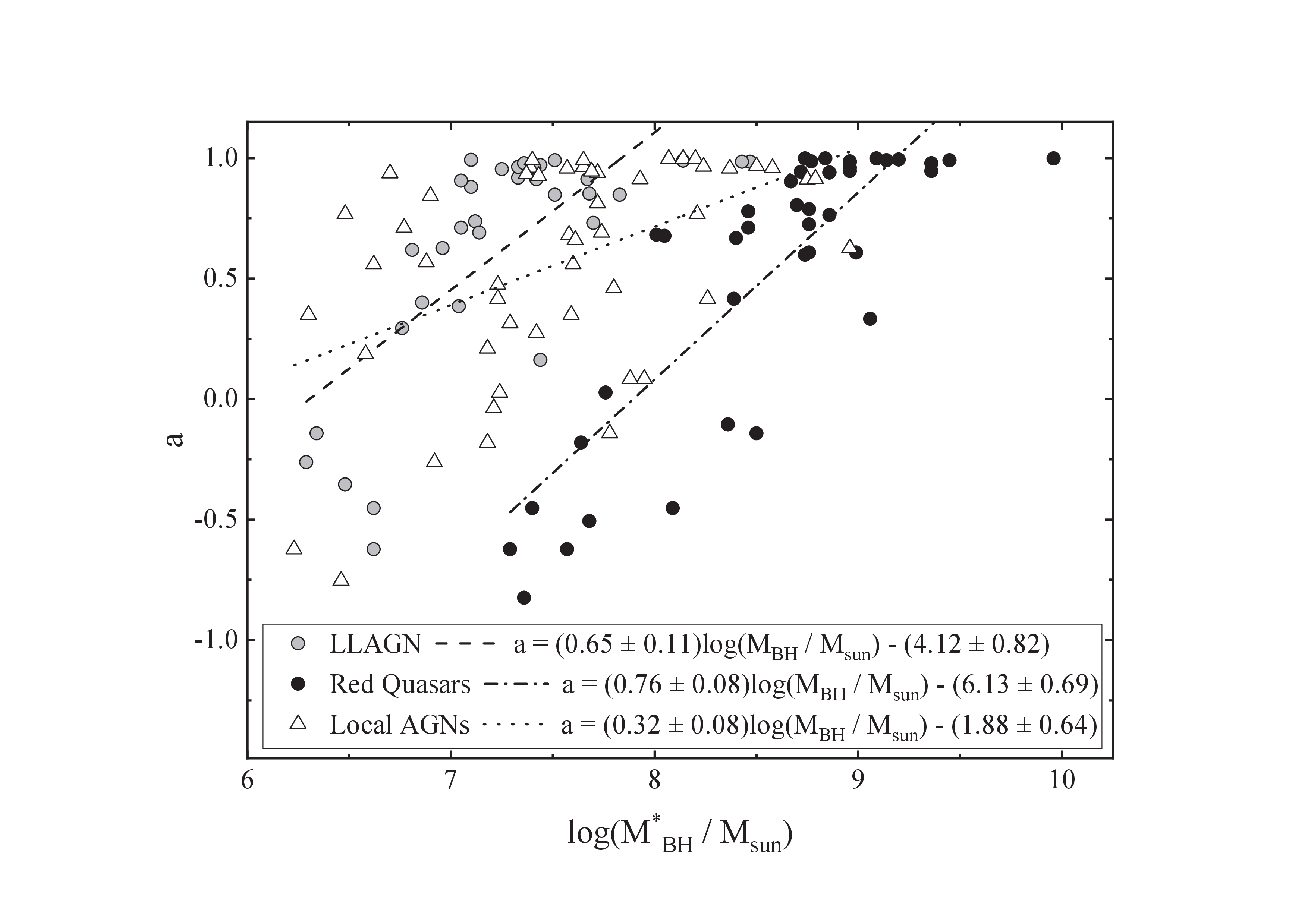}
   \caption{Dependence of the estimated spin values on the estimated SMBH masses for our sample, for sample of red quasars from \citet{piotrovich24} and for sample of local AGNs from \citet{piotrovich22}.}
   \label{M_star-a}
\end{figure}

Fig.~\ref{M_star-a} presents the dependence of the estimated spin values on the estimated SMBH masses. There is strong correlation between them (Pearson correlation coefficient is 0.72). Linear fitting by least squares method gives us $a = (0.65 \pm 0.11)\log(M_\text{BH} / M_\odot) - (4.12 \pm 0.82)$. Thus, we observe a rapid increase in spin with mass, from which we can suggest that the main mechanism of mass growth in this case is disk accretion, which effectively spins up the SMBH. In our work \citet{piotrovich24} we obtain very similar (within error limits) relation for red quasars. It should be noted that for distant ultraluminous AGNs spin increases more slowly with mass \citep{piotrovich25}, however, this is most likely due to the selection bias, as well as the fact that among those objects, a higher percentage have a spin greater than 0.9 (i.e. it could be a saturation effect).


\section{Conclusions}

We estimated radiative efficiency, spin and SMBH mass values for sample of 33 distant LLAGNs.

The dependence of the bolometric luminosity on the SMBH mass for the initial sample shows strong correlation in a form of $\log(L_\text{bol}\text{[erg/s]}) = (0.73 \pm 0.09) \log(M_\text{BH} / M_\odot)) + (39.04 \pm 0.64)$. At the same time a similar dependence for red quasars obtained by us in \citet{piotrovich24} shows a somewhat stronger dependence of luminosity on mass.

The distribution of the estimated spin values (majority of objects have a spin greater than 0.8) is fairly typical for many types of AGNs. Thus we can suggest that our objects in this regard do not have significant differences from most objects of other types.

The dependence of the estimated spin values on the estimated SMBH masses shows strong correlation between them in a form of $a = (0.65 \pm 0.11)\log(M_\text{BH} / M_\odot) - (4.12 \pm 0.82)$. That is, we observe a rapid increase in spin with mass, from which we can suggest that the main mechanism of mass growth in this case is disk accretion, which effectively spins up the SMBH. For example, in our work \citet{piotrovich24} we obtain very similar relation for red quasars. For distant ultraluminous AGNs spin increases more slowly with mass \citep{piotrovich25}, however, this is most likely due to the selection bias, as well as the fact that among those objects, a higher percentage have a spin greater than 0.9 (i.e. it could be a saturation effect).

We did not find any significant qualitative differences in the spin characteristics between our objects and objects of other types considered in the paper. Of course, the results of this study cannot be considered exhaustive. Further studies with a larger number of objects and with more qualitative and extensive observational data are required (in particular, for a more accurate determination of the bolometric luminosity of objects).

\normalem
\begin{acknowledgements}
We are grateful to the Reviewer for very useful comments.
\end{acknowledgements}

\bibliographystyle{raa}
\bibliography{mybibfile}

\end{document}